\documentclass[journal,twocolumn,comsoc]{IEEEtran}

\usepackage{amsmath,amsfonts,bbm,color,comment}
\usepackage[noadjust]{cite}
\usepackage{graphicx}

\newtheorem{theorem}{Theorem}
\newtheorem{lemma}[theorem]{Lemma}
\newtheorem{definition}[theorem]{Definition}

\newcommand{\type}[1]{\emph{#1}}
\newcommand{\etal}{\emph{et al.}}

\renewcommand{\tilde}{\widetilde}

\begin{document}

\title{Individual Testing is Optimal for Nonadaptive Group Testing in the Linear Regime}
\author{Matthew Aldridge
\thanks{The author is with the Department of Mathematical Sciences, University of Bath, Bath, U.K.~and the Heilbronn Institute for Mathematical Research, Bristol, U.K. E-mail: m.aldridge@bath.ac.uk.
%
}}

\maketitle

\begin{abstract}
We consider nonadaptive probabilistic group testing in the linear regime, where each of $n$ items is defective independently with probability $p \in (0,1)$, and $p$ is a constant independent of $n$. We show that testing each item individually is optimal, in the sense that with fewer than $n$ tests the error probability is bounded away from zero.
\end{abstract}

\section{Introduction}

Group testing considers the following problem: Given $n$ items of which some `defective', how many `pooled tests' are required to accurately recover the defective set. Each pooled test is performed on a subset of items: the test is negative if all items in the test are nondefective, and is positive if at least one item in the test is defective.

In Dorfman's original work \cite{dorfman}, the application was to test men enlisting into the U.S.~army for syphilis using a blood test. Dorfman noted that one could test pools of mixed blood samples and use fewer tests than testing each blood sample individually. The test result of such a pool should be negative if every blood sample in the pool is free of the disease, while the test result should be positive if at least one of the blood samples is contaminated. Other applications of group testing include in biology \cite{walter}, signal processing \cite{gilbert},  and communications \cite{varanasi}, to name just a few.

The most important distinction between types of group testing is between:
\begin{itemize}
  \item \type{Nonadaptive testing}, where all the tests are designed in advance.
  \item \type{Adaptive testing}, where the items placed in a test can depend on the results of previous tests.
\end{itemize}
This paper considers nonadaptive testing. Nonadaptive testing is important for modern applications of group testing, where an experimenter wishes to perform a large number of expensive or time-consuming tests which are required to be performed in parallel.

A second important consideration is how many defective items there are. In this paper, we consider the \emph{linear regime}, where the number of defective items $k$ is a constant proportion $p \in (0,1)$ of the $n$ items. A lot of group testing work has concerned the \emph{very sparse regime} $k = O(1)$ \cite{dyachkov-rykov,malyutov,AS} or the \emph{sparse regime} $k = \Theta(n^\alpha)$ for $\alpha < 1$ \cite{BJA,ABJ,SC,KJP,JAS} as $n \to \infty$. However, we argue that the linear regime is more appropriate for many applications. For example, in Dorfman's original set-up, we might expect each person joining the army to have a similar prior probability $p$ of having the disease, and that this probability should remain roughly constant as more people join, rather than tending towards $0$; thus one expects $k \approx pn$ to grow linearly with $n$.

Within nonadaptive group testing in the linear regime, two cases have received most consideration in the literature:
\begin{itemize}
\item \type{Combinatorial zero-error testing:} One wishes to find the defective set with certainty, whichever subset of $\{1,2,\dots,n\}$ of the given size $k$ it is. One assumes that $k = k(n)$ tends to a constant $p \in (0,1)$ as $n \to\infty$. \cite{HHW,RC,huang,aldridge}
\item \type{Probabilistic small-error testing:} We assume each item is defective with probability $p$, independent of all other items, where $p \in (0,1)$ stays fixed. We want to find the defective set with arbitrarily small error probability (averaged over the random defective set). \cite{ungar,mezard,wadayama,AJM,aldridge}
\end{itemize}
This paper considers probabilistic small-error testing. One could consider the case of combinatorial small-error testing, but the probabilistic case is more realistic in applications: again, soldiers might each have a disease with some known prevalence $p$, but it is unrealistic to know \emph{exactly} how many soldiers have the disease. Probabilistic zero-error testing is not of interest: since any of the $2^T$ subsets of items could be the defective set, it is immediate that individual testing is optimal.

The choice between combinatorial (fixed~$k$) or probabilistic (fixed~$p$) set-ups tends not to affect results, as the probabilistic case sees concentration of the number defectives around $k = pn$. The choice between small-error or zero-error can affect the number of tests required in some cases -- for example, nonadaptive testing in the sparse regime, as discussed at the end of this section.

We emphasise that we are looking for full reconstruction; that is, we only succeed if we find the exact defective set, classifying every defective and nondefective item correctly. (See Definition \ref{defs} for formal definitions.)


For group testing in the linear regime, it is easy to see that the optimal scaling is the number of tests $T$ scaling linearly with $n$. A simple counting bound (see, for example, \cite{BJA,KJP,AJM}) shows that we require $T \geq (1-\delta)H(p)n$ for large enough $n$, where $H(p)$ is the binary entropy. Meanwhile, testing each item individually requires $T = n$ tests, and succeeds with certainty. (In the combinatorial case, $T = n-1$ suffices, as the status of the final item can be inferred from whether $k$ or $k-1$ defective items have been already discovered from individual tests.) Thus we are interested in the question: when is individual testing with $T = n$ (or $n-1$) optimal, and when can we reduce $T$ towards the lower bound $H(p)n$?

In the \type{adaptive combinatorial zero-error} case it is known that individual testing is optimal for $p \geq 1 - \log_3 2 \approx 0.369$ \cite{RC} and suboptimal for $p < 1/3$ \cite{HHW,FKW} for all $n$. Hu, Hwang and Wang \cite{HHW} conjecture that $p = 1/3$ is the correct threshold. In forthcoming work, Aldridge \cite{aldridge} gives algorithms using $T < 1.11\,H(p)n$ tests for all $p \leq 1/2$  and large $n$.

In the \type{adaptive probabilistic small-error} case, it is known that individual testing is optimal for $p \geq (3 - \sqrt{5})/2 \approx 0.382$ and suboptimal for $p < (3 - \sqrt{5})/2$ \cite{ungar,aldridge} for $n$ sufficiently large. In forthcoming work, Aldridge \cite{aldridge} gives algorithms using $T < 1.05\,H(p)n$ tests for all $p \leq 1/2$ and large $n$.

In the \type{nonadaptive combinatorial zero-error} case, it is well known that individual testing is optimal when $k$ grows faster than roughly $\sqrt n$, which is the case for all $p \in (0,1)$ in the linear regime for $n$ sufficiently large \cite{dyachkov-rykov,chen-hwang,huang,du-hwang}.

This leaves the \type{nonadaptive probabilistic small-error} case. In this paper, we show that individual testing is optimal for all $p \in (0,1)$ and all $n$.

\begin{theorem} \label{mainthm}
Consider probabilistic nonadaptive group testing where each of $n$ items is independently defective with a given probability $p \in (0,1)$, independent of $n$. Suppose we use $T < n$ tests. Then there exists a constant $\epsilon = \epsilon(p) > 0$, independent of $n$, such that the average error probability is at least $\epsilon$.
\end{theorem}

(The average error probability is defined formally in Definition \ref{defs}.)

The best previously known result was by Agarwal, Jaggi and Mazumdar \cite{AJM}. They used a simple entropy argument to show that individual testing is optimal for $p \geq (3 - \sqrt 5)/2 \approx 0.382$, and a more complicated argument using Madiman--Tetali inequalities to extend this to  $p > 0.347$. We extend this to all $p \in (0,1)$. Further, Agarwal \etal\ use a weaker definition of `optimality' than we do here: they show that the error probability is bounded away from $0$ as $n \to \infty$ when $T < (1-\delta)n$ for some $\delta > 0$, whereas we show that the error probability is bounded away from $0$ for any $T < n$.

Wadayama \cite{wadayama} had claimed to be able to beat individual testing for some $p$ in work that was later retracted in part \cite{wadayama2}. We discuss this matter further in Section \ref{three}.

Finally, we note that other scaling regimes than the linear regime have been studied, notably the sparse regime $k = \Theta (n^\alpha)$ for different values of the sparsity parameter $\alpha \in [0,1)$. In these regimes, for $n$ sufficiently large: adaptive testing always outperforms individual testing \cite{hwang,huang,BJA}, nonadaptive small-error testing always outperforms individual testing \cite{AS,chan,ABJ,SC,JAS}, and nonadaptive zero-error testing outperforms individual testing for $\alpha < 1/2$ but not for $\alpha > 1/2$ \cite{dyachkov-rykov,chen-hwang,KS,du-hwang}.

\section{Definitions and notation}

We fix some notation and recap some important definitions.

\begin{definition} \label{defs}
There are $n$ items, and we perform $T$ tests. A nonadaptive test design can be defined by a test matrix $\mathsf X = (x_{ti}) \in \{0,1\}^{T\times n}$, where $x_{ti} = 1$ means item $i$ is in test $t$, and $x_{ti} = 0$ means it is not.

Given a test design $\mathsf X$ and a defective set $\mathcal K \subseteq \{1,2,\dots,n\}$, the outcomes $\mathbf y = (y_t) \in \{0,1\}^T$ are given by $y_t = 0$ if $x_{ti} = 0$ for all $i \in \mathcal K$, and $y_t = 1$ otherwise.

An estimate of the defective set is a (possibly random) function $\hat{\mathcal K} = \hat{\mathcal K}(\mathsf X, \mathbf y) \subseteq \{1,2,\dots,n\}$.

The \emph{average error probability} is
  \[ \mathbb P(\mathrm{error}) = \sum_{\mathcal K \subseteq \{1,2,\dots,n\}} p^{|\mathcal K|} (1-p)^{n - |\mathcal K|}  \, \mathbb P \big(\hat{\mathcal K}(\mathsf X, \mathbf y) \neq \mathcal K\big) ,\]
where $\mathbf y$ is related to $\mathsf X$ and $\mathcal K$ as above, and the probability $\mathbb P$ can be replaced by an indicator function if the estimate $\hat{\mathcal K}$ is nonrandom.
\end{definition}

The concept of an item being `disguised' will be important later.

\begin{definition} \label{disg}
Fix a test design $\mathsf X$ and a defective set $\mathcal K$. Given an item $i$ (either defective or nondefective) contained in a test $t$, we say that item $i$ is \emph{disguised} in test $t$ if at least one of the other items in that test is defective; that is, if there exists a $j \in \mathcal K, j\neq i$, with $x_{tj}=1$. We say that item $i$ is \emph{totally disguised} if it is disguised in every test it is contained in.
\end{definition}

\begin{lemma} \label{lemma}
Consider probabilistic group testing with defective probability $p$. Fix a test design $\mathsf X$, 
and write $w_t = \sum_{i=1}^n x_{ti}$ for the \emph{weight} of test $t$; that is, the number of items in test $t$. Further, write $D_i$ for the event that item $i$ is totally disguised. Then
  \begin{equation*} \label{one}
    \mathbb P(D_i) \geq \prod_{t : x_{ti} = 1} (1 - q^{w_t-1}) \, ,
  \end{equation*}
where $q = 1-p$.
\end{lemma}

This is essentially Lemma 4 of \cite{AJM}; we give a shorter proof here based on the FKG inequality (see for example \cite{FKG}, \cite[Section 2.2]{Grimmett}).

\begin{IEEEproof}
For a test $t$ containing item $i$, write $D_{t,i}$ for the event that $i$ is disguised in $t$. Clearly we have
\[
  \mathbb P(D_i) = \mathbb P \left(\, \bigcap_{t : x_{ti} = 1} D_{t,i} \right)  .
\]
Further, for a test $t$ containing item $i$ we have $\mathbb P(D_{t,i}) = 1-q^{w_t - 1}$, since $i$ is disguised in $t$ unless the other $w_t - 1$ items in the test are all nondefective.
Note also that the $D_{t,i}$ are \emph{increasing events}, in the sense that for $\mathcal L \subseteq \mathcal K$ the indicator functions satisfy $\mathbbm 1_{D_{t,i}}(\mathcal L) \leq \mathbbm 1_{D_{t,i}}(\mathcal K)$. The FKG inequality tells us that increasing events are positively correlated, in that
  \[ \mathbb P \left(\, \bigcap_{t : x_{ti} = 1} D_{t,i} \right) \geq \prod_{t : x_{ti} = 1} \mathbb P(D_{t,i}) \, , \]
and the result follows.
\end{IEEEproof}

\section{Proof of main theorem}

We are ready to proceed with the proof of Theorem \ref{mainthm}. 

\begin{IEEEproof}[Proof of Theorem \ref{mainthm}]
The key idea is the following: Suppose some item $i$ is totally disguised, in the sense of Definition \ref{disg}. Then every test containing $i$ is positive, no matter whether $i$ is defective or nondefective. Thus we cannot know whether $i$ is defective or not: we either guess $i$ is nondefective and are correct with probability $p$, guess $i$ is nondefective and are correct with probability $q$, or take a random choice between the two. Whichever way, the error probability is bounded below by the constant $\min\{p,q\}$, which is nonzero for $p \in (0,1)$. It remains to show that, again with probability bounded away from $0$, there is such a totally disguised item $i$.

Fix $n$. Fix a test design $\mathsf X$ with $T < n$ tests. 
Without loss of generality we may assume there are no tests of weights $w_t=0$ or $1$. All weight-$0$ `empty' tests can be removed. If there is a weight-$1$ test, we can remove it and the item it tests, repeating until there are no weight-$1$ tests remaining. These removals leave $p$ the same, do not increase the error probability, and reduce $T/n$, since we had $T/n < 1$ to start with.

From Lemma \ref{lemma}, the probability that item $i$ is totally disguised is bounded by 
  \begin{equation*} 
    \mathbb P(D_i) \geq \prod_{t : x_{ti} = 1} (1 - q^{w_t-1}) \, ,
  \end{equation*}
Write $L(i)$ for the logarithm of this bound, so $\mathbb P(D_i) \geq \mathrm e^{L(i)}$, where
  \begin{align*}
    L(i) &= \log \prod_{t : x_{ti} = 1} (1 - q^{w_t-1}) \\
      &= \sum_{t : x_{ti} = 1} \log(1 - q^{w_t-1}) \\
      & = \sum_{t=1}^T x_{ti} \log(1 - q^{w_t-1}) .
  \end{align*}
We must show that, for some $i$, $L(i)$ is bounded from below, independent of $n$. Then with probability at least $\mathrm{e}^{L(i)}$ we have a totally disguised item, and the theorem follows.

Write $\bar L$ for the mean value of $L(i)$, averaged over all $i$ items. (Note that $\bar L$ is negative.) Then we have
\begin{align}
  \bar L &= \frac1n \sum_{i=1}^n L(i) \notag \\
    &= \frac1n \sum_{i=1}^n \sum_{t=1}^T x_{ti} \log(1 - q^{w_t-1}) \notag \\
    &= \frac1n \sum_{t=1}^T \left( \sum_{i=1}^n x_{ti}\right) \log(1 - q^{w_t-1}) \notag \\
    &= \frac1n \sum_{t=1}^T w_t \log(1 - q^{w_t-1}) \notag \\
    &\geq \frac{T}{n} \,\min_{t=1,2,\dots,T} \left\{w_t \log(1 - q^{w_t-1}) \right\} \label{here} \\
    &\geq \min_{t=1,2,\dots,T} \left\{w_t \log(1 - q^{w_t-1}) \right\} \label{there} \\
    &\geq \min_{w=2,3,\dots} \left\{w \log(1 - q^{w-1}) \right\}  =: L^* . \label{Lstar}
\end{align}
Going from \eqref{here} to \eqref{there} we have used the assumption $T/n \leq 1$ (and that $T/n$ is multiplied by a negative expression), and going from \eqref{there} to \eqref{Lstar} we used assumption that no test has weight $0$ or $1$. Note further that the the bound $L^*$ is indeed finite, since, for $q < 1$, the function $w \mapsto w \log (1-q^{w-1})$ is continuous for $w \in [2,\infty)$, finite at $w =2$, and tends to $0$ as $w \to \infty$.

Since $\bar L$ is the mean of the $L(i)$s, there is certainly some $i$ with $L(i) \geq \bar L \geq L^*$, and thus some $i$ with $\mathbb P(D_i) \geq \mathrm e^{L^*}$. We are done. 
\end{IEEEproof}

\begin{figure}
\centering
\includegraphics[width=0.47\textwidth]{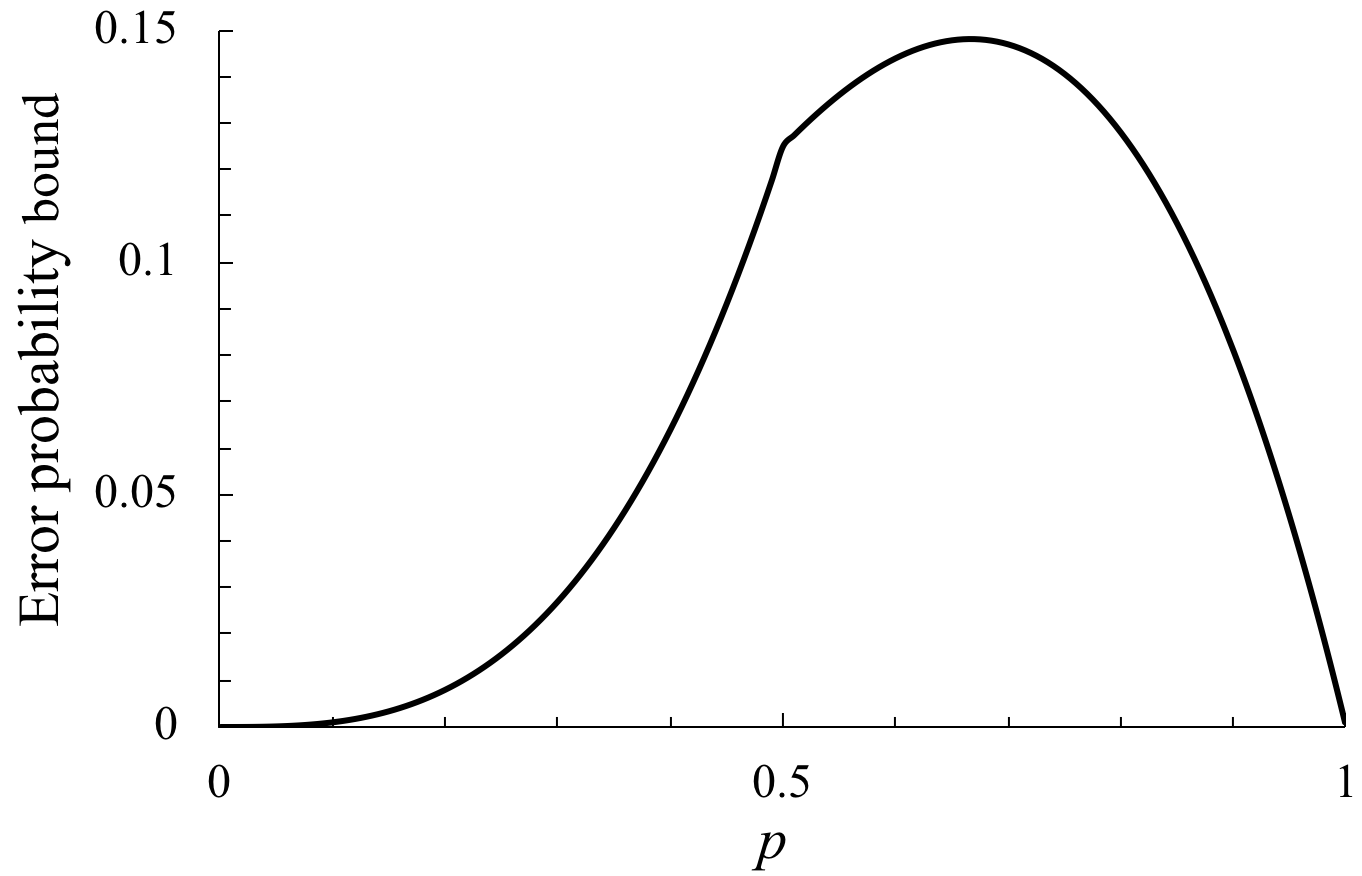}
\caption{The error probability bound \eqref{errprob1}, plotted against the prevalence $p$.}
\label{fig}
\end{figure}

Inspecting the proof, we see immediately that, when $T < n$, we have an explicit bound on the error probability of
  \begin{equation} \label{errprob1}
    \mathbb P(\mathrm{error}) \geq \epsilon(p) = \min\{p,q\} \mathrm e^{L^*} ,
  \end{equation}
with $L^*$ as in \eqref{Lstar}. This is very easy to compute given $p$. The bound \eqref{errprob1} is plotted in Figure \ref{fig}. By being more careful at the step from \eqref{here} to \eqref{there}, we see that when $T < (1-\delta)n$ we can improve the bound \eqref{errprob1} to 
  \begin{equation*} \label{errprob2}
    \mathbb P(\mathrm{error}) \geq \min\{p,q\} \mathrm e^{(1-\delta)L^*} . 
  \end{equation*}

Note that in the proof we this only bounded the probability that one particular item is wrongly decoded, so the bound, while explicit, simple to compute, and bounded away from $0$, is unlikely to be tight.

\section{Closing remarks} \label{three}

We have shown that individual testing is optimal for small-error nonadaptive testing in the linear regime for all $p \in (0,1)$.

Our result here contradicts a result of Wadayama \cite[Theorem 2]{wadayama}, later retracted \cite{wadayama2}, which claimed arbitrarily low error probabilities for $n$ sufficiently large with $T/n < 1$. Wadayama used \emph{doubly regular} test designs chosen at random subject to each item being in $l$ tests and each test containing $r$ items, where $l$ and $r$ are kept fixed as $n \to \infty$. Since $l/r = T/n$, one requires cases with $r > l \geq 1$ to beat individual testing. However, with these designs, following the outline above, we see that the probability that any given item $i$ is totally disguised is bounded below by $\mathbb P(D_i) \geq (1-q^{r-1})^l$, a constant greater than $0$ for $r > l \geq 1$. Thus these designs cannot have arbitrarily low error probability.
  
A note of caution with our result is due. We have only shown that the error probability cannot be made \emph{arbitrarily} small -- however, it might be very small.  For example, we see from Figure \ref{fig} that the error bound given by \eqref{errprob1} is very small for $p < 0.1$.  Thus, for some given nonzero error tolerances, and some $p$ and $n$, it may still be that random designs can be profitably used in applications -- suggestions include Bernoulli random designs \cite{malyutov,AS,chan,ABJ,SC,mezard}, designs with constant tests-per-item \cite{JAS,mezard}, or Wadayama's doubly regular designs \cite{wadayama,mezard}. Further `finite blocklength' analysis of these designs would be useful in investigating this point.

We have shown that individual testing is optimal in the linear regime $k \sim pn$ for $p>0$, while it is known that individual testing is suboptimal when $k = \Theta(n^\alpha)$ for any $\alpha < 1$ \cite{AS,chan,ABJ,SC}. This leaves open exactly when individual testing becomes suboptimal. For example, is individual testing optimal or not when $k \sim n/\log n$? The method employed here required $p$ to be bounded away from $0$; with $p = 1/\log n$, for example, a totally disguised item could be safely assumed to be nondefective with error probability tending to $0$.




\IEEEtriggeratref{13}

\bibliographystyle{IEEEtran}
\bibliography{bibliography}

\end{document}